\documentclass[prb,
superscriptaddress,showpacs,amsmath,amssymb]{revtex4}

\usepackage{graphicx}

\begin{document}

\title{Dirac and Weyl fermions: from Gor'kov equations to Standard Model\\
(in memory of Lev Petrovich Gorkov)}

\author{G.E.~Volovik}
\affiliation{Low Temperature Laboratory, Department of Applied Physics, Aalto University, PO Box 15100, FI-00076 AALTO, Finland}
\affiliation{Landau Institute for Theoretical Physics RAS, Kosygina 2, 119334 Moscow, Russia}
\affiliation{P.N. Lebedev Physical Institute, RAS, Moscow 119991, Russia}

\date{\today}

\begin{abstract}
Gor'kov theory of superconductivity \cite{Gorkov1958} opened the application of the methods of quantum field theory to condensed matter physics. Later the results became relevant to relativistic quantum fields.
\end{abstract}

\maketitle

 \section{Introduction}

Application of quantum field theory to condensed matter physics  began in Soviet Union  around 1956-57
\cite{Gorkov2011}. In this approach the Fermi sea serves as an analog of the relativistic quantum vacuum -- the Dirac sea. The Gor'kov theory of superconductivity  \cite{Gorkov1958} has been the fundamental step in this direction, which in turn  triggered the development of the relativisic theories. The
composite models developed by Nambu and Jona-Lasinio  \cite{Nambu1961} 
and by Vaks and Larkin  \cite{VaksLarkin1961}, where the Higgs bosons appear as a composite states of the fermion pairs, are the direct consequences of the Gor'kov theory. 
In such models the original Weyl fermions of Standard Model (such as top quarks) play the role of the electrons in metals, while the composite Higgs bosons are analogs of the collective modes of the order parameter in superconductors.

\begin{figure}[top]
\centerline{\includegraphics[width=0.4\linewidth]{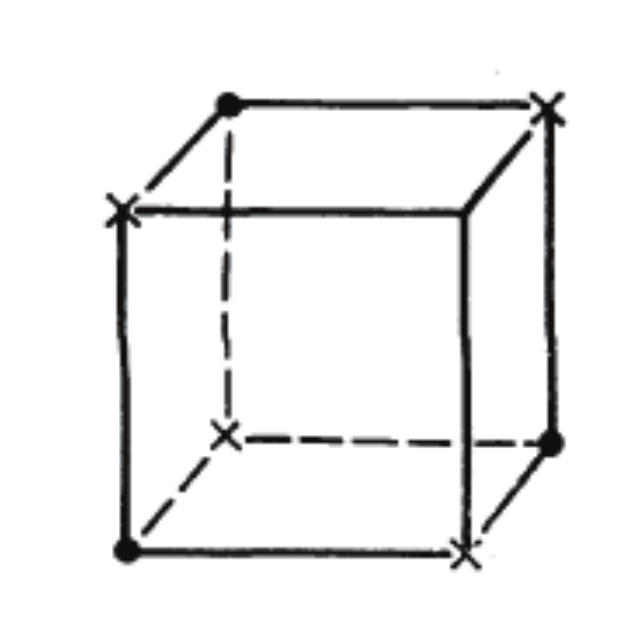}}
\label{Cube} 
  \caption{Figure from Ref. \cite{VolovikGorkov1985}. Arrangement of the nodes in the energy spectrum in superconductors of class $O(D_2)$. The points denote
four Weyl nodes  with topological charge $N=+1$, and crosses denote four Weyl nodes  with $N=-1$. 
The modern expression for the topological charge in terms of the Gor'kov-Nambu Green'sfunction is in Eq.(\ref{N3}).
In the vicinity of each Weyl node with $N=+1$ the chiral right-handed Weyl fermions emerge, while $N=-1$ is the topological charge of the left-handed quasiparticles. This can be compared with 8 right-handed and 8 left-handed particles (quarks and leptons) in each generation of Standard Model fermions.
 }
\end{figure}

Here we consider another consequence of the Gor'kov theory of superconductivity, where the Weyl fermions emerge in superconductors as Bogoliubov quasiparticles. This in particular takes place for superconductors of the symmetry class $O(D_2)$, where the  4 left-handed and 4 right-handed topologically protected chiral fermions emerge \cite{VolovikGorkov1984,VolovikGorkov1985}, see Fig. \ref{Cube}.  Expansion of the Gor'kov Green's function in the vicinity of each topologically protected Weyl point leads to the effective relativistic quantum field theory with effective gauge fields and the effective gravity. This provides the hint for possible emergent origin of the "fundamental" Weyl fermions, gauge fields, and general relativity \cite{FrogNielBook,Horava2005,Volovik2003}.

%%%%%%%%%%%%%%%%%%%%%%%%%%%%%%%%%%%%%%%%
%%%%%%%%%%%%%%%%%%%%%%%%%%%%%%%%%%%%%%%%
%%%%%%%%%%%%%%%%%%%%%%%%%%%%%%%%%%%%%%%%
%%%%%%%%%%%%%%%%%%%%%%%%%%%%%%%%%%%%%%%%

\section{Symmetry and topology}
\label{SymmetryTopology}

The Gor'kov anomalous function $F$ and the ordinary Green's function $G$ have been organized by Nambu \cite{Nambu1960} in terms of the $2\times 2$  matrix Hamiltonian, and Gor'kov equations have been written in the more convinient matrix form. 
The corresponding $2\times 2$ Green's function is  
\begin{equation}
  \hat G^{-1}=i\omega +   \tau_3\epsilon({\bf p}) + 
\tau_1{\it Re}\,\Delta({\bf p})  + \tau_2{\it Im}\,\Delta({\bf p}) \,.
 \label{GorkovGreen}
 \end{equation}
Here $\Delta({\bf p})$ is the complex gap function; $\tau_{1,2,3}$ are $2\times 2$ Pauli matrices, they operate in the particle-hole space and describe the effective spin emerging in the superconducting state -- the Bogoliubov-Nambu-Gor'kov spin. This is in addition to the original spin of electrons, which we ignore here at the moment assuming for simplicity the spin singlet state of the Cooper pair.

The symmetry classification of superconducting states \cite{VolovikGorkov1984,VolovikGorkov1985} demonstrates that the majority of supercoductivity classes contains the gapless superconductors, where the symmetry supports the point nodes and/or the nodal lines in the energy spectrum $E({\bf p})= \sqrt{\epsilon^2({\bf p}) +|\Delta({\bf p})|^2}$ of Bogoliubov quasiparticles, for review see \cite{Gorkov1987}. Some of the nodal lines and points do not disappear even if the symmetry of the superconducting state is violated: such stability towards deformations of crystal is supported by the integer-valued topological invariant, which for the point nodes has been discussed in Ref. \cite{VolovikGorkov1985}. This invariant reflects the topology of the Gor'kov-Nambu matrix Green's function
 in the 3D momentum space or in the 4D momentum-frequency space \cite{GrinevichVolovik1988}. 

Since the topologically protected nodes in the energy spectrum determine the dynamics, kinetics,   thermodynamics and anomalies in superconductors,  the symmetry classification of superconducting states should be supplemented by the 
topological classification. The topological classification of nodes in the energy spectrum with and without the symmetry protection has been started \cite{Horava2005}, but still is not completed, see the latest reviews on symmetry and topology of superconducting and superfuid states \cite{Mizushima2016,Sato2016}.

\section{Weyl and Dirac superconductors}
\label{WeylDiracSuperconductor}

\subsection{Weyl superconductors}
\label{WeylSuperconductor}

In order to make connection to the vacuum of quantum field theory, we  consider the superconducting state in the crystals with the maximal possible symmetry -- the cubic symmetry (note that as has been found by Gor'kov \cite{Gorkov1984},  even in the cubic crystals the upper critical field $H_{c2}$ can be anisotropic). The state, where the gap nodes obtained from the symmetry consideration appear to be topologically stable, belongs to the superconducting class $O(D_2)$. This state contains 8 point nodes in the energy spectrum in Fig. \ref{Cube}, which are situated on the four $C_3$ symmetry axes in the Brillouin zone (the more general situation with the pockets of the Fermi surfaces see in Ref. \cite{Agterberg1999}). Let us consider the gap function $\Delta({\bf p})$ for the momentum ${\bf p}$ along one of these threefold axes.
The important property of the unconventional superconductivity discussed in   \cite{VolovikGorkov1984,VolovikGorkov1985}, is that the symmetry group of the order parameter (or of the gap function $\Delta({\bf p})$) contains the elements with combined symmetry. In particular,
the symmetry group of the $O(D_2)$ state contains the combined symmetry: 
$\tilde C_3= C_3\exp\left(\frac{2\pi i}{3}\right)$. This combined symmetry means that the gap function does not change if the $C_3$ rotation of the momentum ${\bf p}$ around one of the threefold axis is accompanied by the $U(1)$ gauge transformation by the phase $\pi/3$.
Applying this to the gap function $\Delta({\bf p})$ with momentum along the $C_3$ rotation axis one obtains \cite{VolovikGorkov1985}:
\begin{equation}
\Delta({\bf p})=\tilde C_3\Delta({\bf p})= C_3e^\frac{2\pi i}{3}\Delta({\bf p})=
e^\frac{2\pi i}{3}\Delta(C_3{\bf p})= e^\frac{2\pi i}{3}\Delta({\bf p})\,.
\label{C3}
\end{equation}
This gives $\Delta({\bf p})=0$ for the momentum ${\bf p}$ on the threefold axis.

This can be visualized using the
simple model Hamiltonian for fermions  in the  spin-singlet $d$-wave superconductor in cubic crystal, which belongs to $O(D_2)$ symmetry class. The gap function has the form
 \begin{equation}
\Delta({\bf p})= a(p_x^2 +\epsilon p_y^2 +\epsilon^2 p_z^2) \,\,,
 \,\,\epsilon=\exp\left(\frac{2\pi i}{3} \right)\,.
\label{OD2}
\end{equation}
The corresponding Gor'kov-Nambu Hamiltonian for fermions is
 \begin{equation}
 H=\epsilon({\bf p})\tau_3 +  a\frac{1}{2}(2p_x^2 -p_y^2-p_z^2)\tau_1 + a\frac{\sqrt{3}}{2}(p_y^2-p_z^2)\tau_2 \,.
\label{Hamiltonian}
\end{equation}
The intersections of the lines of zeroes in the gap function, $\Delta({\bf p})=0$, with the Fermi surface, $\epsilon({\bf p})=0$, give the point nodes in the energy spectrum. In the considered case of cubic symmetry of the crystal there are  8 point nodes at the vertices of cube in the momentum space in Fig. \ref{Cube}:
 \begin{equation}
{\bf p}^{(n)}=\frac{p_F}{\sqrt{3}}( \pm \hat{\bf x} \pm \hat{\bf y}   \pm \hat{\bf z} )~~,~~n=1, \ldots,8\,.
\label{Nodes}
\end{equation}

It appears that these zeroes in the energy spectrum do not disappear if the crystal is deformed and the cubic symmetry is violated. The reason for that is that these point nodes are topologically stable: they are supported by topological invariant, which takes values $N=\pm 1$ \cite{VolovikGorkov1985}. This invariant is related to the winding number of the phase of the gap function $\Delta({\bf p})=|\Delta({\bf p})|e^{i\Phi({\bf p})}$ around the nodal line in the gap function \cite{VolovikMineev1982}. It also represents the charge of the Berry phase monopoles in momentum space \cite{Volovik1987}, 
and can be analytically expressed in terms of the Gor'kov-Green's function \cite{GrinevichVolovik1988}: 
\begin{equation}
N= \frac{e_{\alpha\beta\mu\nu}}{24\pi^2} \,
{\bf tr}\left[ \int_\sigma   dS^\alpha
 \hat G\partial_{p_\beta}  \hat G^{-1}
 \hat G\partial_{p_\mu}  \hat G^{-1}  \hat G\partial_{p_\nu}   \hat G^{-1}\right].
\label{N3}
\end{equation}
Here $\sigma$ is the 3D surface around  the point nodes of the inverse Green's function in the 4D  frequency-momentum space
$p_\mu=(\omega,{\bf p})$; and ${\bf tr}$ denotes the trace over all the indices of the matrix  $\hat G(p_\mu)$ (in addition to Bogoliubov-Nambu-Gor'kov spin index, the Green's function contains also the ordinary spin index, band index in crystals, and indices of the fermionic species in particle physics). 
In a different form the topological stability of the Weyl nodes has been discussed for the neutrino sector in Refs. \cite{NielsenNinomiya1981a,NielsenNinomiya1981b}.

The important property of the point nodes with $N=\pm 1$ is that close to each of  8 nodes the Gor'kov-Nambu Hamiltonian is reduced to the Hamiltonian describing the Weyl fermions:
\begin{equation}
 H^{(n)}=
e^{{(n)}i}_\alpha \tau^\alpha \left(p_i -p_i^{(n)}  \right) \,\, ,\,\, n=1, \ldots ,8\,.
\label{Weyl}
\end{equation}
Under space and time dependent deformations the expansion parameters  $e^{{(n)}i}_\alpha({\bf r},t)$  become the fields. These are analogous to the tetrad field in general relativity acting on Weyl fermion. In the same way 
$p_i^{(n)}({\bf r},t)$ play the role of the effective gauge fields.

Altogether the $O(D_2)$ superconducting state contains 4 left-handed fermions ($N=-1$)  and 4 right-handed fermions ($N=+1$). In case of Fermi pockets discussed in Ref. \cite{Agterberg1999} the number of Weyl fermions may increase. If the pockets are on the three-fold axes, one would have 8 left-handed and 8 right-handed fermions.
This can be compared with 8 left-handed and 8 right-handed Weyl fermions in each generation of fermions in Standard Model of particle physics. The natural arrangement of the 8 left and 8 right Weyl fermions also occurs on the vertices of the cube in the four-dimensional $(\omega,p_x,p_y,p_z)$ space \cite{Creutz2008,Creutz2014}.
This is one of many examples when the topologically protected nodes in the spectrum serve as an inspiration
for the construction of the relativisitc quantum field theories.

%%%%%%%%%%%%%%%%%%%%%%%%%%%%%%%%%%%%%%%%
%%%%%%%%%%%%%%%%%%%%%%%%%%%%%%%%%%%%%%%%
\begin{figure}[top]
\centerline{\includegraphics[width=0.4\linewidth]{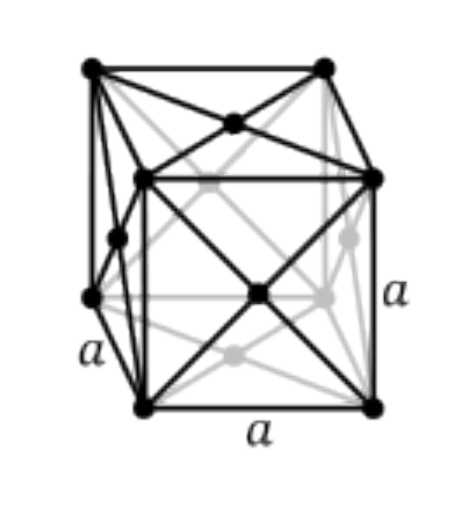}}
\label{Cube2} 
  \caption{Unstable point nodes in the triplet state of the symmetry class $O(T)$.
  This state has $14=8+6$ point nodes \cite{VolovikGorkov1985}. The zeroes are at the 8 points, where the Fermi surface intersects the four threefold axes, and at 6 points at the intersections of the Fermi surface with three fourfold axes. These points may split into pairs of Weyl points if the time reversal symmetry is violated. The extension to the 4D space $(\omega,p_x,p_y,p_z)$ may give $24=16+8$ point nodes, and up to 48 Weyl fermions, if the time reversal symmetry is violated.
 }
\end{figure}
%%%%%%%%%%%%%%%%%%%%%%%%%%%%%%%%%%%%%%%%
%%%%%%%%%%%%%%%%%%%%%%%%%%%%%%%%%%%%%%%%

\subsection{Dirac superconductors}
\label{DiracSuperconductor}

Topological stability of nodes in the spectrum is important, because it does not depend on the microscopic mechanism of superconductivity. On the other hand, the symmetry arguments are rather subtle,
especially in case of complicated many-band structure of the Fermi surface and of the gap function, and also the symmetry can be violated.
An example of state with unstable Fermi points is the triplet state of the symmetry class $O(T)$ with Gor'kov-Nambu Green's function 
\begin{equation}
  \hat G^{-1}=i\omega +   \tau_3\epsilon({\bf p}) + 
\tau_1\sigma_i d_i({\bf p}) \,,
 \label{Triplet}
 \end{equation}
where the vector gap function is
\begin{equation}
 {\bf d}({\bf p}) \propto \hat{\bf x}p_x(p_y^2-p_z^2) + \hat{\bf y}p_y(p_z^2-p_x^2)+ \hat{\bf z}p_z(p_x^2-p_y^2)  \,.
\label{vectord}
\end{equation}
Here $\sigma_i$ are Pauli matrices for ordinary spin, and the energy spectrum $E({\bf p})= \sqrt{\epsilon^2({\bf p}) +{\bf d}^2({\bf p})}$ in case of real vector ${\bf d}({\bf p})$.

This state has $14=8+6$ point nodes \cite{VolovikGorkov1985}. The zeroes are at the 8 points, where the Fermi surface intersects the four threefold axes, and in addition there are 6 points at the intersections of the Fermi surface with three fourfold axes, see Fig. \ref{Cube2}. This state is time reversal invariant (the vector $ {\bf d}({\bf p})$ is real), and thus the nodes do not represent the Weyl points. These are the Dirac points, which are massless (gapless) only due the symmetry of the state, and can be destroyed by deformations of the crystal lattice. Depending on the type of the violated symmetry, the fermions may acquire the gap (mass), or instead the Dirac points split into the pairs of Weyl points 
\cite{KlinkhamerVolovik2004,KlinkhamerVolovik2005}. The first scenario takes place in Standard Model, with the possible exception in the neutrino sector, where the second scenario is still not excluded.

The extension to the four-dimensional $(\omega,p_x,p_y,p_z)$ space with cubic symmetry gives 
$16+8=24$ Dirac points, which can be compared with 24 Dirac fermions in 3 generations of Standard Model.
The crystal symmetry of superconductors does not allow us to consider the point groups of  higher symmetry, such as the icosahedral group and its 4D extensions.  However, this can be extended by considering the symmetry groups in the internal isospin space \cite{Lampe2013,Lampe2015}.

\section{Cuprates and Dirac nodal lines}
\label{lines}

Though the mechanism of high temperature superconductivity (HTSC) in cuprates remains
unknown \cite{Gorkov2008}, the $d$-wave order parameter symmetry and existence of the nodal lines are well established.
Let us start with cuprates which have tetragonal symmetry.  The superconducting state there belongs to the class $D_4(D_2)$.  The relevant combined symmetry, which gives rise to the nodes in the spectrum,  is $U_2e^{\pi i}$. It is the $\pi$ rotation about the horizontal axis combined with the $U(1)$ gauge transformation by phase $\pi/2$ \cite{VolovikGorkov1985}. Under this combined transformation the gap function 
\begin{eqnarray}
\Delta(p_x,p_y,p_z)=U_2e^{\pi i}\Delta(p_x,p_y,p_z)=-\Delta(p_y,p_x,-p_z)=
\nonumber
\\
-\Delta(-p_y,-p_x,-p_z)=-\Delta(p_y,p_x,p_z)\,.
\label{U2}
\end{eqnarray} 
Here we used the symmetry element $C_2$ of the group $D_4(D_2)$ in the third equality and the space inversion symmetry in the last one.
The equation (\ref{U2})  gives $\Delta({\bf p})=0$ for momenta on the vertical plane $p_x = p_y$. The same consideration shows that the gap function is zero also on another vertical plane, $p_x = -p_y$. As a result  one obtains 4 nodal lines in the energy spectrum at the intersections
of the Fermi surface $\epsilon({\bf p})=0$ with the two vertical symmetry planes, where   $\Delta({\bf p})=0$.
The corresponding Gor'kov-Nambu  model Hamiltonian is
 \begin{equation}
 H=\epsilon({\bf p})\tau_3 +  a(p_x^2 -p_y^2)\tau_1 \,.
\label{HamiltonianCuprate}
\end{equation}
These nodal lines survive if the tetragonal symmetry is distorted, but the time reversal symmetry is obeyed.
The reason for that is again the topological stability.
The nodal lines are protected by another topological invariant expressed via the general 
 Gor'kov-Nambu  Hamiltonian
  \begin{equation} 
N= {1\over 4\pi i} ~{\rm tr} ~\oint_C dl ~ \tau_2 H^{-1}\nabla_l H \,.
\label{InvariantForLine}
\end{equation}
Here the integral is now along the loop $C$ around the nodal line in 3D momentum space.

Recently the polar phase of superfluid $^3$He has been discovered, which has the spin degenerate nodal line in the energy spectrum \cite{Dmitriev2015}.

\section{Physical consequences of gap nodes in superconductors}

 The nodes in the spectrum lead to the finite density of states (DoS) in the presence of the superfluid velocity, this is because the nodes expand due to the Doppler shift and form the Fermi surface pockets around the former nodes.  Due to the superflow around Abrikosov vortices, the Fermi pockets produce the singular dependence of the DoS on magnetic field.  The point nodes lead to the DoS, which is proportional to ${\bf v}_s^2$. Integration over the cross section of the Abrikosov vortices gives the $B\ln B$ dependence of DoS  in the vortex lattice \cite{Volovik1988}. In case of the line nodes the DoS has linear dependence on $|{\bf v}_s|$, see Ref. \cite{MuzikarRainer1983}, and correspondingly one obtains  the $\sqrt{B}$ dependence of DoS in the Abrikosov lattice \cite{Volovik1988}. This square root dependence of DoS has been observed in cuprates \cite{Moler1994}.

The systems with Weyl fermions should experience the chiral anomaly effect \cite{Adler,BellJackiw} in the external fields or in the effective gauge fields produced by deformations. The latter has been observed in chiral superfluid $^3$He-A as an extra force acting on vortex-skyrmions \cite{Bevan1997}. This is the spectral flow force or the Kopnin  force, see e.g. Ref. \cite{Kopnin1995}. It can be expressed in terms of the effective "magnetic" ${\bf B}$  and "electric" ${\bf E}$  fields acting on Weyl fermions. These fields are produced by deformation in superconductors, or by the moving skyrmion texture in $^3$He-A \cite{Volovik2003}. The Kopnin force is proportional to ${\bf B}\cdot {\bf E}$, and the measured coefficient in front of ${\bf B}\cdot {\bf E}$ was found to be in agreement with the Adler-Bell-Jackiw equation \cite{Adler,BellJackiw}, desribing production of the fermionic charge generated by the chiral anomaly.  

The topologicaly protected point and line nodes in the energy spectrum lead to the fermion zero modes in the core of vortices and to the topologically protected gapless edge states. This is the result of the so-called bulk-surface and bulk-vortex topological correspondence. The Weyl fermions in bulk produce the dispersionless flat band of fermions with zero energy living in the vortex core \cite{Kopnin1991,Volovik2011} and the Fermi arc of the edge states \cite{Burkov2011,Burkov2011b}. 
The nodal lines in bulk superconductors lead to the condensation of the Caroli-de Gennes-Matricon levels in the vortex \cite{Volovik2016} and to the flat band of the edge states \cite{Ryu2002,SchnyderRyu2011}.

The  dimensional reduction of the 3D superconducting states with Weyl fermions to the 2D or quasi-3D 
films leads to the two-dimensional fully gapped superconductivity without the gap nodes, which however retains the nontrivial topological character. The topological invariant for these 2D superconductors is obtained by dimensional reduction of Eq.(\ref{N3}) for the Weyl point. It is expressed in terms of Gor'kov-Nambu matrix Green's function, but the integration now is over the frequency (on the imaginary axis) and over the 2D Brillouin zone,
see Refs. \cite{So1985,IshikawaMatsuyama1986,IshikawaMatsuyama1987,Volovik1988b}:
\begin{equation}
N= \frac{e_{\beta\mu\nu}}{24\pi^2} \,
{\bf tr}\left[ \int d\omega d^2p\,
 \hat G\partial_{p_\beta}  \hat G^{-1}
 \hat G\partial_{p_\mu}  \hat G^{-1}  \hat G\partial_{p_\nu}   \hat G^{-1}\right].
\label{2D}
\end{equation}
Here $p_\mu=(p_x,p_y,\omega)$.
The  Abrikosov vortex in such 2D superconductors contains in its core the Majorana fermion with zero energy \cite{Volovik1999,ReadGreen2000,Ivanov2001}.

  \section{Conclusion}

 The majority of the topological classes of superconductivity contain the topologically protected Weyl points.
Expansion of the Gor'kov-Nambu matrix Green's function in the vicinity of each Weyl point leads to the effective relativistic quantum field theory with effective gauge fields and effective gravity. In this analogy different Weyl points correspond to different fermionic species of Standard Model - quarks and lepton.  The effective quantum field theory in the vicinity of the nodal line
is still waiting for its development, as well as in the vicinity of nodes of arbitrary co-dimension. Probably this may have relation to branes in string theory \cite{Horava2005}.

\section*{\hspace*{-4.5mm}ACKNOWLEDGMENTS}
 The support by RSCF (No. 16-42-01100) is acknowledged.

 \end{document}